# Collimating lenses from non-Euclidean transformation optics

**Kan Yao[1,2,3], Xunya Jiang[2,3] and Huanyang Chen[1,3]**

[1] School of Physical Science and Technology, Soochow University, Suzhou, Jiangsu 215006, China

[2] The State Key Laboratory of Functional Materials for Informatics, Shanghai Institute of Microsystem and Information Technology, Chinese Academy of Sciences, Shanghai 200050, China

[3] E-mail: ustcykk@gmail.com, xyjiang@mail.sim.ac.cn and chy@suda.edu.cn

**Abstract.** Based on the non-Euclidean transformation optics, we design a thin metamaterial lens that can achieve wide-beam radiation by embedding a simple source (a point source in three-dimensional case or a line current source in two-dimensional case). The scheme is performed on a layer-by-layer geometry to convert curved surfaces in virtual space to flat sheets, which pile up and form the entire lens in physical space. Compared to previous designs, the lens has no extreme material parameters. Simulation results confirm its functionality.

---

[3] Author to whom any correspondence should be addressed.

# 1. Introduction

Achievements in metamaterials in the past decade have enabled unconventional control of electromagnetic fields, such as perfect imaging [1] and directive emission [2]. More recently, by reaching the ultimate optical illusion—invisibility cloak [3, 4], transformation optics attracted great attention and soon became an active issue. In simple terms, transformation optics utilizes the form-invariance property of Maxwell's equations, which ensures a spatial coordinate transformation can be interpreted as (or equivalent to) the effect of a substitution of media. This fact provides a bright way to manipulate electromagnetic fields on both wave and ray scales, facilitating the design of new materials and advanced devices [5]. Besides the fascinating prospect in perfect imaging and invisibility, many practical instruments, for example, lenses to collimate, bend, split or shift beams have also been proposed as the applications of metamaterials. It is demonstrated that a point source embedded in a slab of near-zero index material radiates energy only in a narrow cone [2]. Later, the usage of non-Euclidean transformation optics inspires more solutions to this subject. By setting radius-dependent branch cuts, a spherical antenna can focus light into a needle-sharp beam [6]. Conventional transformation method is particularly straightforward when producing wide beams with high directivity, since arbitrary wavefronts can be assigned to flat ones [7]-[11]. However, the design of metamaterial lenses seems to face a natural dilemma, that when the lens is tuned to be relatively thin (compared to the width of lens or beam), the material parameters always require extreme values [2, 6-10] and sometimes a wide source [9], while the devices will become thick if the singularities are removed [11].

In this paper, we discuss an idea to achieve directive radiation via a thin lens made of transformation media. The transformation, which is based on a three-dimensional extension of the stereographic projection on two-dimensional manifolds, maps the wavefronts within a sphere gradually into a flat sheet, and limits the thickness of the device to a quarter of its width. Meanwhile, as no singular point is involved, all material parameters avoid extreme values. Holding the same geometry in cross section, this scheme should also be valid for two-dimensional case, and numerical simulations confirm this deduction.

## 2. Non-Euclidean Transformation

We begin by introducing two bases that are necessary to present the detailed idea. Firstly, the mathematical tool is visualized by a conformal transformation, the stereographic projection. Invented by Ptolemy for cartographic use, amazingly, the projection also maps the non-Euclidean metric of a spherical surface to the index profile of Maxwell's fisheye in flat space [12, 13]. Figure 1(a) shows in cross section how the projection works, where a plane is inserted through the equator of a sphere. By connecting a straight line to the north pole, points $(x, y)$ on the plane establish a one-to-one correspondence with points $(X, Y, Z)$ on the surface of the sphere,

$$R = \sqrt{X^2 + Y^2 + Z^2}, \quad x = \frac{X}{1 - Z/R}, \quad y = \frac{Y}{1 - Z/R}. \tag{1}$$

This convenience allows elegant operations on spheres to be expressed on planes, where it might be difficult to imagine, e.g., setting non-Euclidean branch cuts [6, 14].

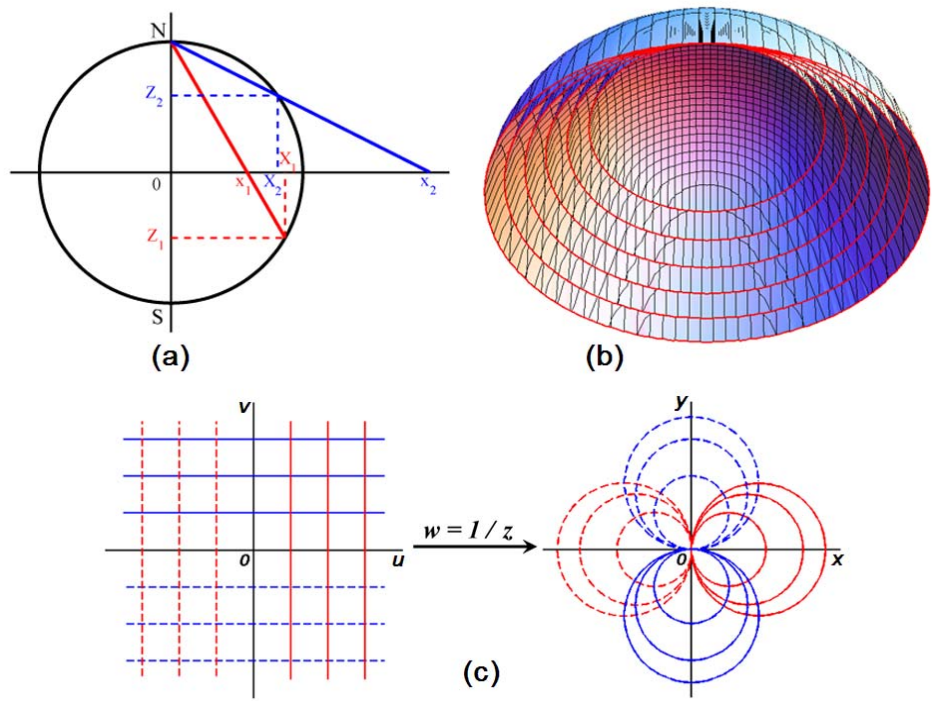

**Figure 1.** Basic formulae and geometry. (a) Stereographic projection maps each point $(X, Y, Z)$ on the surface of the sphere to a projected point $(x, y)$ on the plane and vice versa. The southern hemisphere is mapped to the disk bounded by the equator, and the northern hemisphere is mapped beyond. (b) The three-dimensional space inside a sphere is presented by a set of spherical layers tangent at the north

pole. (c) A complex inversion proves in cross section that how the geometry in (b) covers the entire volume.

The second basis, namely the geometrical tool to be used later is an unusual representation of the three-dimensional space. Although it is natural to describe the volume inside a sphere as a foliage of concentric spherical layers [6], here we adopt a different style that all the Riemann spheres share a common north pole and are tangent to each other there, as illustrated in figure 1(b). Can this geometry cover the entire space? The answer is yes. To show this, consider two complex planes denoted by $w = u + \mathrm{i}v$ and $z = x + \mathrm{i}y$, respectively. With a Möbius transformation

$$w = \frac{1}{z} = \frac{x - \mathrm{i}y}{x^2 + y^2}, \tag{2}$$

while $u$-axis or $v$-axis sweeps over an arbitrary half-plane in $w$ space, their images, a set of circles through the origin in $z$ space, shrink continuously from infinity to zero in the corresponding half-plane, see figure 1(c). The proof for three-dimensional case can be obtained by simply rotating this two-dimensional geometry about either axis.

Now we present the idea for designing the lens. To achieve high-directive and wide-beam radiation, the wavefront on the aperture of the device should be flat. From the viewpoint of non-Euclidean transformation optics, a spherical wavefront radiated by a point source could be mapped to a flat one by stereographic projection. We thus set the boundary of the lens in virtual space—the surface of a sphere with fixed radius, and all the transformations should be performed within it. It is reasonable to imagine that the region of transformation in virtual space is filled with a set of concentric spherical wavefronts, however, we cannot map these layers into flat sheets like the boundary. The cause is straightforward. When the wavefronts shrink to the source, their curvatures tend to infinity; hence the media are not able to flatten them without extreme values [6, 9]. To remove this singularity, we adopt an unusual style to represent the region of transformation, as shown in figure 2. The interior of the boundary is covered by a set of tangent Riemann spheres, which are fixed at a common north pole *N*. Suppose we perform the stereographic projection on the southern hemisphere of each layer, a corresponding set of disks will naturally pile up like Hanoi Tower. The transformation

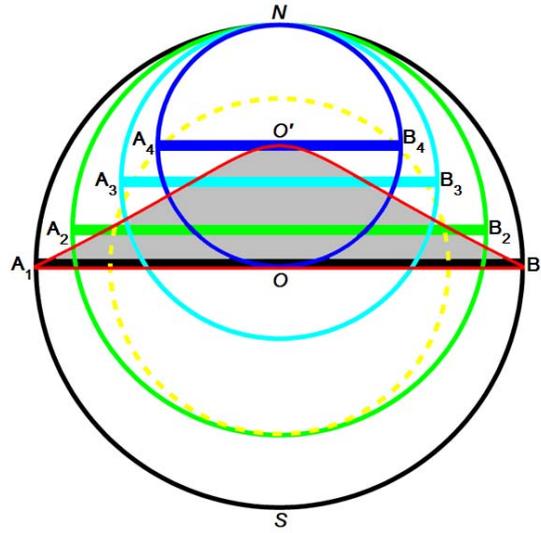

**Figure 2.** Schematic of the transformation. Stereographic projection is performed on each layer of the tangent Riemann spheres. Since the north pole $N$ is fixed, spheres shrink to $N$ rather than the origin $O$, and the corresponding planes of projection rise synchronously to fit the locations of spheres. Each southern hemisphere is mapped to a disk inside the equator (indicated by the sequence of heavy lines $A_nB_n$ for $n = 1, 2, 3, ....$ ) These disks naturally pile up and occupy a three-dimensional volume. Particularly, the space below $A_1B_1$ that describes half of the radiation pattern of a simple source at $O$ is mapped to the gray region bounded by red curves, i.e. the lens, and the source is projected to $O'$ without causing any singularity in material.

is given by

$$x = \frac{X}{R_0 - Z} R, \qquad y = \frac{Y}{R_0 - Z} R, \qquad z = R_0 - R \tag{3}$$

with

$$R = \frac{X^2 + Y^2 + (R_0 - Z)^2}{2(R_0 - Z)}. \tag{4}$$

Here $R$ and $R_0$ denote the radii of inner Riemann spheres and outer boundary, respectively. Only the volume below equator $A_1B_1$ ($Z \leq 0$) is needed for projection.

The mapping is not an assignment of wavefronts except on the boundary, since the Riemann spheres do not coincide with equiphase surfaces. However, this geometry removes singularity from the origin to the north pole, which is not involved in our consideration. Noting that the curvature of each wavefront layer is larger than that of the corresponding Riemann sphere, like, e.g., the yellow dashed line and green solid line in figure 2, the wavefronts in physical space become more and more warped from the aperture to the source, and from a flat disk to a single point. Another advantage is about the size. From figure 2, it is not difficult to conclude that the thickness of the lens is limited to one quarter of the width.

## 3. Discussion

A necessary procedure before further discussion is to calculate the material parameters. Using the formulae given in (3) and (4), one can obtain the tensor of permittivity and permeability as

$$\bar{\bar{\varepsilon}} = \bar{\bar{\mu}} = \frac{\Lambda \Lambda^T}{\det(\Lambda)} = 8 \cdot [1 + \frac{x^2 + y^2}{(R_0 - z)^2}]^{-1} \begin{pmatrix} \frac{1}{4} + \frac{x^2}{(R_0 - z)^2} & \frac{xy}{(R_0 - z)^2} & -\frac{1}{2}\frac{x}{R_0 - z} \\ \frac{xy}{(R_0 - z)^2} & \frac{1}{4} + \frac{y^2}{(R_0 - z)^2} & -\frac{1}{2}\frac{y}{R_0 - z} \\ -\frac{1}{2}\frac{x}{R_0 - z} & -\frac{1}{2}\frac{y}{R_0 - z} & \frac{1}{4} \end{pmatrix}, \quad (5)$$

where $\Lambda$ is the Jacobian matrix of the transformation. Noticeably, the material is anisotropic. The reason is that although the stereographic projection maps each Riemann sphere to a flat sheet conformally, its extension is not conformal when performed layer-by-layer to cover a three-dimensional volume [6, 15]. Unlike in isotropic media, the trajectory of a light ray in anisotropic materials normally does not coincide with the direction of wave vector. Therefore, while a hemispherical wavefront is converted into a flat disk continuously, the light rays act different, see figure 3. Inside the anisotropic lens, light rays that emitted from a point source at vertex travel smoothly and intersect the aperture at different angles. Once passing through the interface and entering free space, they are suddenly confined to a uniform direction along the wave vector, which is always normal to the wavefronts. Seen from three-dimensional view, rays within a light cone experience refraction at the interface and then emerge as a beam with high-directivity.

The constitutive tensor in equation (5) is given by a complicated expression. To get more information and check the feasibility for fabricating, we derive the eigenvalues of the tensor so that it can be written into a diagonal form. Denoting the eigenbasis with ($u$, $v$, $w$), the three components are

$$\varepsilon_u = \mu_u = \frac{4}{1+\Delta}(\frac{1}{2}+\Delta+\sqrt{\Delta^2+\Delta}), \quad \varepsilon_v = \mu_v = \frac{4}{1+\Delta}(\frac{1}{2}+\Delta-\sqrt{\Delta^2+\Delta}), \quad \varepsilon_w = \mu_w = \frac{2}{1+\Delta},$$

(6)

where $\Delta = (x^2 + y^2)/(R_0 - z)^2$. The third eigenvalue is particularly worthy of note. For a given height $z = z_0$, namely a certain sheet parallel to the aperture, the spatial dependence of $\varepsilon_w$ and $\mu_w$ follows the refractive index profile of Maxwell's fisheye. This result is reasonable since we perform the transformation by mapping a set of tangent Riemann spheres layer-by-layer to flat sheets via stereographic projection, which normally converts uniform spherical surfaces into Maxwell's fisheyes on planes [12, 13]. Although for each infinitesimally thin layer, Maxwell's fisheye only deals with light rays on it, when the layers pile up, interestingly, the volume contributes to the control of light propagating with the third dimension. Figure 4 shows the three eigenvalues in cross section. As seen, $\varepsilon_w$ and $\mu_w$ range from 1 to 2 and vary as Maxwell's fisheye does on each horizontal layer, where the minimum value is tailored by the height-dependent outline. Meanwhile, as a result of avoiding the singular point in transformation, the other two eigenvalues are also limited in a finite range.

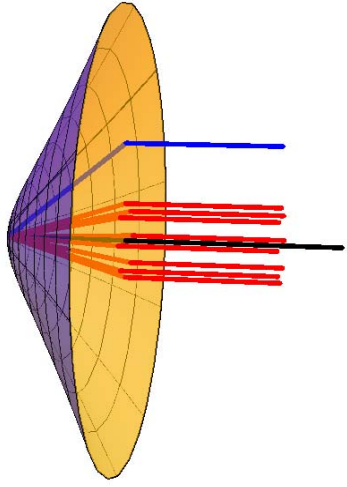

**Figure 3.** Light collimating and focusing in three-dimensional view. Via the lens, a light cone emitted from the point source at vertex is bent to a bundle of light rays (shown in red) that normal to the aperture plane. The focusing can be displayed conversely. Both effects work on all the light rays enter this device, including the non-paraxial ones (shown in blue).

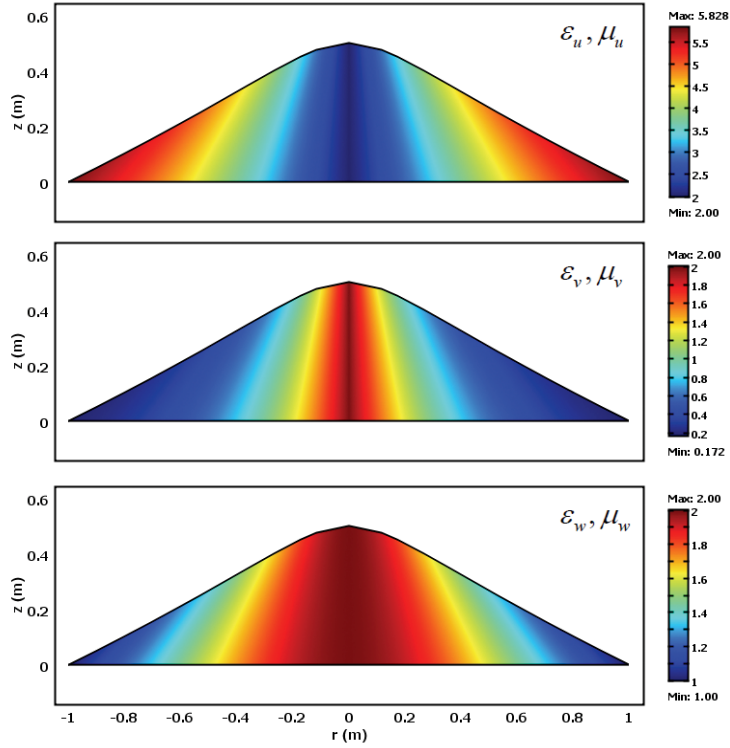

**Figure 4.** The material parameters of the lens expressed in eigenbasis ($u$, $v$, $w$) by choosing $R_0 = 1\text{m}$ in equation (6).

The formula in equation (3) leads to a three-dimensional device; however, sharing the same geometry in cross section, one can repeat similar procedure in two-dimensional space, in which the stereographic projection conformally maps a circle into a line. In this case, the lens acts as a cylindrical-to-plane-wave converter [8]. We omit the detailed derivation and presentation of material parameters here but directly show the numerical results. In simulations, a line current source is located a little inside the vertex, and the outline is covered by a perfect electric conductor (PEC) shell to confine the radiation. Figure 5 gives the electric field distribution at 2.5 GHz and 6.5 GHz. The different behaviors of wavefronts and light rays (denoted by power flow lines in gray) can be observed in both cases: the wavefronts are converted gradually and smoothly, while the light rays suffer refraction at the interface. Once leaving the aperture, light rays coincide well with the normal direction of wavefronts, which means a high-directive emission.

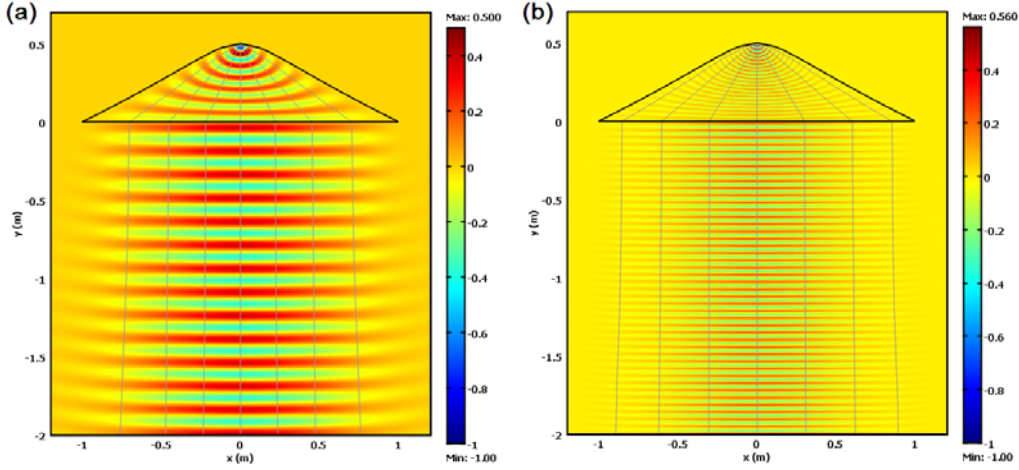

**Figure 5.** Electric field distribution due to a line current source located at the vertex of the two-dimensional lens at (a) 2.5 GHz and (b) 6.5 GHz. The power flow lines (in gray) as well as the equiphase contours show how light rays and wavefronts propagate, respectively.

At last, we would like to discuss the possibility for further reducing the thickness of the lens. This question is equivalent to achieve a larger width-to-thickness ratio. Actually, the factor is determined by the geometrical representation of the virtual space, see figure 6(a) for instance. The space is still presented by a set of tangent spheres. Using a different definition of stereographic projection that settles the plane at the south pole, one can obtain a wider lens (in red) compared with the original scheme (in blue). But unfortunately, the translation of projected plane is a linear operation; thus the thickness and width are enlarged in the same scale. To mainly squeeze the thickness, we need to change the manner of describing the space. Figure 6(b) shows an alternative style with bipolar coordinates. Composing an orthogonal coordinate system, the curved axes, which illustrate four sets of circles, can cover the entire plane completely. Furthermore, one can rotate bipolar coordinates about *y*-axis or *x*-axis to present the three-dimensional space with sets of spheres in toroidal coordinates or bispherical coordinates. Both options are available to perform the layer-by-layer transformation like equation (3), but noting the difference how the spheres intersect, only the bispherical coordinates can be used to obtain a thinner lens. Put it into mathematics, we have

$$x = a \cdot \coth\tau, \qquad y = \frac{Y}{1 - \frac{X - a \cdot \coth\tau}{a/\sinh\tau}}, \qquad z = \frac{Z}{1 - \frac{X - a \cdot \coth\tau}{a/\sinh\tau}} \qquad (7)$$

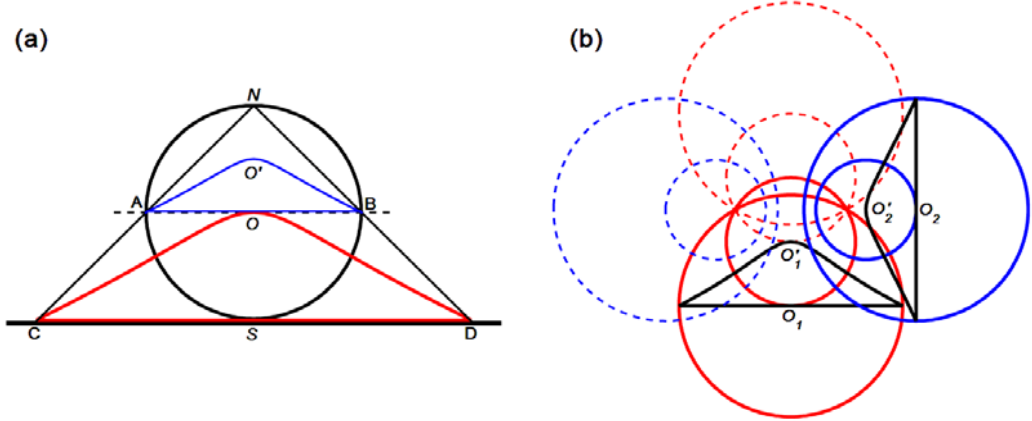

**Figure 6.** Lens flattening. (a) Translation of planes in the definition of stereographic projection leads to lenses in different scales, but the geometry is not changed. (b) By using bipolar coordinates to describe the virtual space, we can adjust the ratio of width to thickness. With red circles intersecting at the two foci, the lens is thicker than the original design in figure 2, since the circle through source $O_1$ is larger than the one tangent at the north pole and therefore is mapped to a higher position $O_1'$. Conversely, if choose the blue circles surrounding the foci, the resulting lens will be thinner.

with

$$(X - a \cdot \coth \tau)^2 + Y^2 + Z^2 = a^2 / \sinh^2 \tau, \tag{8}$$

where $a$ denotes the foci and $\tau$ is the variable in bispherical coordinates describing the position and radius of each sphere. Compared with figure 2, the scenario here can make the lens very thin in theory. The cost of this improvement is, the derivation will be much more complicated as the spheres do not share a common north pole for projection, and, the maximum and minimum of the first two eigenvalues will increase and decrease respectively (but still finite), while the third eigenvalue keeps its spatial dependence like Maxwell's fisheye.

## 4. Conclusion

In summary, we present an idea to design a thin lens that converts spherical waves to a collimated beam. The lens is made of transformation media. By performing stereographic projection layer-by-layer to a set of tangent Riemann spheres, the transformation avoids singular points; hence the material parameters vary in a finite range. Both light-ray model and numerical results show good performance on the directivity of emission. Possible improvement on thickness is discussed by changing the representation of the space, and more interesting devices could be expected by acting different operations on the layer-by-layer geometries.

**Acknowledgments**

This work was supported by the National Natural Science Foundation of China (grant nos. 11004147, 60877067 and 11174309), the Natural Science Foundation of Jiangsu Province (grant no. BK2010211) and the Priority Academic Program Development (PAPD) of Jiangsu Higher Education Institutions.